\documentclass{appolb}
\usepackage{epsfig,color}
\usepackage{amssymb}
\begin{document}
\pagestyle{plain}
\newcount\eLiNe\eLiNe=\inputlineno\advance\eLiNe by -1
\title{Correlation matrix decomposition of WIG20 intraday fluctuations
}
\author{R.Rak$^a$, S.~Dro\.zd\.z$^{a,b}$,
J.~Kwapie\'n$^b$, P.~O\'swi\c ecimka$^b$
\address{$^a$Institute of Physics, University of Rzesz\'ow, PL--35-310
Rzesz\'ow, Poland\\
$^b$Institute of Nuclear Physics, Polish Academy of Sciences, \\
PL--31-342 Krak\'ow, Poland}} \maketitle

\begin{abstract}

Using the correlation matrix formalism we study the temporal
aspects of the Warsaw Stock Market evolution as represented by the
WIG20 index. The high frequency (1 min) WIG20 recordings over the
time period between January 2001 and October 2005 are used. The
entries of the correlation matrix considered here connect
different distinct periods of the stock market dynamics, like days
or weeks. Such a methodology allows to decompose the price
fluctuations into the orthogonal eigensignals that quantify
different modes of the underlying dynamics. The magnitudes of the
corresponding eigenvalues reflect the strengths of such modes. One
observation made in this paper is that strength of the daily trend
in the WIG20 dynamics systematically decreases when going from
2001 to 2005. Another is that large events in the return
fluctuations are primarily associated with a few most collective
eigensignals.
\end{abstract}


\section{Introduction}

Nature of the temporal correlations in financial fluctuations
constitutes one of the most fascinating issues of the contemporary
physics. The pure Brownian-type motion~\cite{Bachelier} is
definitely not an optimal reference~\cite{Mandelbrot}. Already the
correlations in the financial volatility remain positive over a
very long time horizon~\cite{Mantegna}. Even more involved are
higher order correlations that give rise to the financial
multifractality~\cite{Oswiecimka}. In this context a mention
should also be given to the concept of financial log-periodicity -
a phenomenon analogous to criticality in the discrete scale
invariant version~\cite{Sornette, Drozdz1}.

One more approach, initiated in Ref.~\cite{Kwapien1}, to quantify
the character of financial time-correlations~\cite{Kwapien2} is to
use a variant of the correlation matrix. In this approach the
entries of the corresponding matrix connect the high-frequency
time series of returns representing different disconnected
time-intervals like the consecutive days or weeks. The structure
of eigenvectors of such a matrix allows then to quantify several
characteristics of time correlations that remain unvisible by more
conventional methods.

\begin{figure}[h!]
\hspace{0.5cm} \epsfxsize 11.5cm \epsffile{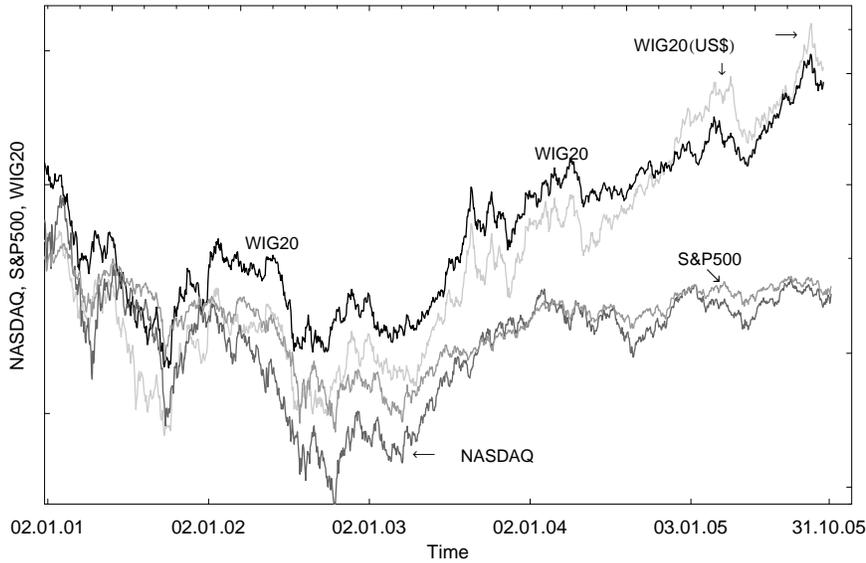} \caption{The
WIG20 (Warsaw Stock Exchange), the S$\&$P500 and the Nasdaq
indices from 2001.01.02 until 2005.10.31.}
\end{figure}

Using this methodology here we present a systematic study for the
Polish stock market index WIG20 over the period
02.01.2001-31.10.2005. The corresponding WIG20 chart, expressed
both in terms of the Polish Zloty (PLN) and in terms of the
US$\$$, versus two world leading stock market indices: the Nasdaq
and the S$\&$P500, is shown in Fig.~1. The Warsaw Stock Exchange
trading time during this period was 10:00-16:10 and the WIG20
recorded with the frequency of 1 min.

\section{Formalism}

In the present study  the correlation matrix is thus defined as
follows. To each element in a certain sequence $N$ of relatively
long consecutive time-intervals of equal length $K$ labeled with
$\alpha$ one uniquely assigns a time series $x_{\alpha}(t_i)$,
where $t_i$ $(i=1,2,...,K)$ is to be understood as discrete time
counted from the beginning for each $\alpha$. In the financial
application $x_{\alpha}(t_i)$ is going to represent the price
time-series, $\alpha$ the consecutive trading days (or weeks) and
$t_i$ the trading time during the day (week). As usual it is then
natural to define the returns $R_{\alpha}(t_i)$ time-series as
$R_{\alpha}(t_i) = \ln x_{\alpha}(t_i+\tau) - \ln
x_{\alpha}(t_i),$ where $\tau$ is the time lag. The normalized
returns are defined by
\begin{equation}
r_{\alpha}(t_i) = {R_{\alpha}(t_i) - \langle R_{\alpha}(t_i)
\rangle_t \over v} \label{eq2}
\end{equation}
where $v$ is the standard deviation of returns over the period $T$
and $ v^2 = \sigma^2(R_{\alpha}) = \langle R_{\alpha}^2(t)
\rangle_t - \langle R_{\alpha}(t) \rangle_t^2\ , $ and
$\langle\ldots\rangle_t$ denotes averaging over time.\\
One thus obtains $N$ time series $r_{\alpha}(t_i)$ $(\alpha =
1,...,N)$ of length $T$=$K$-1, {\it i.e.} an $N \times T$ matrix
$\bf M$. Then, the correlation matrix is defined as $ {\bf C} = (1
/ T) \ {\bf M} {\bf M}^{\bf T}$. By diagonalizing $\bf C$
\begin{equation}
{\bf C} {\bf v}^k = \lambda_k {\bf v}^k, \label{eq4}
\end{equation}
one obtains the eigenvalues $\lambda_k$ $(k=1,...,N)$ and the
corresponding eigenvectors ${\bf v}^k = \{ v^k_{\alpha} \}$. In
the limiting case of entirely random correlations the density of
eigenvalues $\rho_C(\lambda)$ is known
analytically~\cite{Pastur,Edelman,Sengupta}, and reads \\
$~~~~~~~~~~~~~~~~~~~~~~~~~~~~~\rho_C(\lambda) = {Q \over {2 \pi
\sigma^2}} {\sqrt{ (\lambda_{max} - \lambda) (\lambda -
\lambda_{min})} \over {\lambda}},$

\begin{equation}
\lambda^{max}_{min} = \sigma^2 (1 + 1/Q \pm 2 \sqrt{1/Q}),
\label{eq6}
\end{equation}

with $\lambda_{min} \le \lambda \le \lambda_{max}$, $Q=T/N \ge 1$,
and where $\sigma^2$ is equal to the variance of the time series
(unity in our case).\\
For a better visualization, each eigenvector can be associated
with the corresponding time series of returns by the following
expression:
\begin{equation}
z_k(t_i) = \sum_{\alpha=1}^N v_{\alpha}^{k} r_{\alpha}(t_i),
~~~~~~k = 1,...,N; ~~~ i=1,...,T. \label{eq7}
\end{equation}
These new time series thus decompose the return fluctuations into
the orthogonal components that reflect distinct patterns of
oscillations common to all the time intervals labeled with
$\alpha$. They are therefore called the
eigensignals~\cite{Kwapien1,Kwapien2}.

\section{Results}
\subsection{Correlations among trading days}

The above methodology is here applied to the WIG20 1 min
recordings during the period between January 02, 2001 and October
31, 2005. This whole time period is split and analysed separately
for the consecutive calendar years $\mathcal{Y}$ that it covers
($\mathcal{Y}=\{2001,2002,2003,2004,2005\}$). The number
$N_{\mathcal{Y}}$ of trading days correspondingly equals 249, 243,
249, 255, and 210. The WIG20 intraday variation is systematically
taken between the trading time 10:01:30 (at this time the index is
always already determined) and 16:10:00. This corresponds to
$T=368$ during one trading day. Using these data sets we construct
the $N_{\mathcal Y}\times N_{\mathcal Y}$ correlation matrices
$\bf C_{\mathcal{Y}}$.

\begin{figure}[h!]
\hspace{-1.6cm} \epsfxsize 16.0cm \epsffile{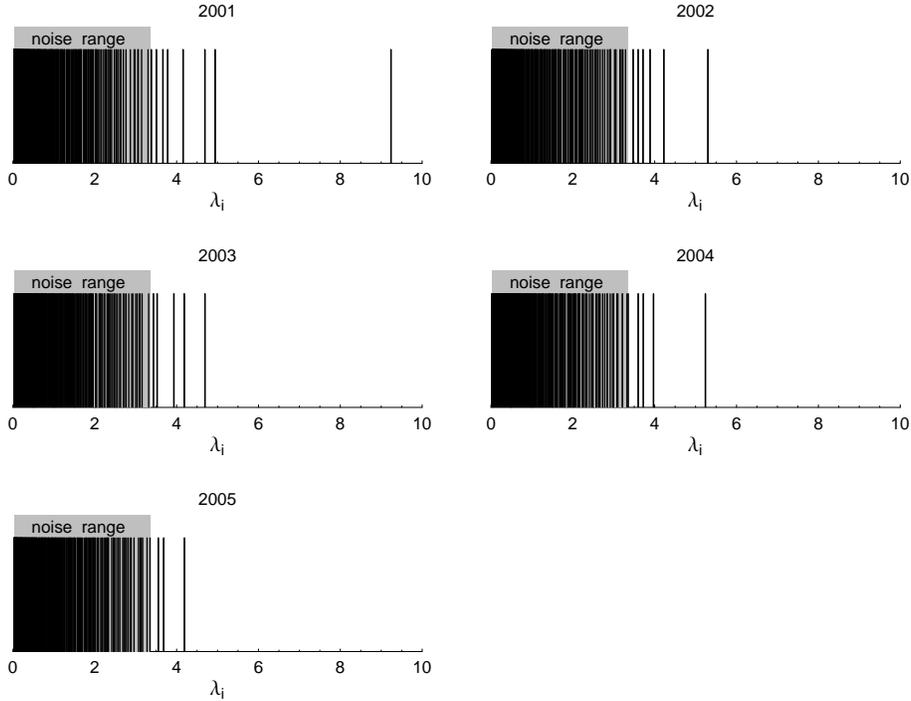}
\caption{Empirical eigenvalue spectrum of the correlations
matrices $\bf C_{\mathcal{Y}}$ (vertical black lines) calculated
for WIG20(Warsaw Stock Exchange) index over the five consecutive
calendar years. The noise range, as determined by a random Wishart
matrix with $Q= 368/N_{\mathcal{Y}}$, is indicated by the shaded
field.}
\end{figure}

The structure of eigenspectrum $\lambda_k^{\mathcal{Y}}$ of such
matrices for all the five calendar years is shown in Fig.~2. The
pure noise range - as prescribed (Eq.~(\ref{eq6})) by the
corresponding Wishart ensemble of random
matices~\cite{Edelman,Sengupta} - is indicated by the shaded area.
As one can see, the majority of eigenvalues of our empirical
correlation matrices are located within this area which signals
that noise is dominating. Typically there exist however several
eigenvalues that stay significantly above it. They are associated
with some collectivity effects that in the present case are to be
interpreted as an appearance of certain repeatable structures in
the intraday dynamics of financial fluctuations. Definitely one
such structure is the daily trend. As far as the WIG20 dynamics is
concerned it is however even more interesting to see that when
going from 2001 to 2005 those large eigenvalues gradually decrease
and get closer to the noise area. This effect is more
systematically documented in Fig.~3 which shows the evolution of
the four largest eigenvalues of the $N \times N=250\times250$
correlation matrix, which corresponds to 250 consecutive trading
days, and this time window is moved with a step of one month (20
trading days).

The structure of eigenspectrum is expected to be closely related
to the distribution of matrix elements of the correlation
matrices. For the same five calendar years as in Fig.~2 the
corresponding distributions are shown in Fig.~4 versus their
Gaussian best fits. Indeed, in 2001 this distribution deviates
most from a Gaussian and even develops a power-law tail with the
slope of $\gamma \approx 6$ ($P(x)\sim x^{-\gamma}$). It is in
this case that the largest eigenvalue of the correlation matrix is
repelled most (upper most left panel of Fig.~2) from the rest of
the spectrum due to an effective reduction of the rank of the
matrix~\cite{Drozdz2}. Later on the distribution of the matrix
elements is much better fit by a Gaussian and some deviations
remain on the level of essentially single entries.

\begin{figure}[h!]
\hspace{0.0cm} \epsfxsize 12cm \epsffile{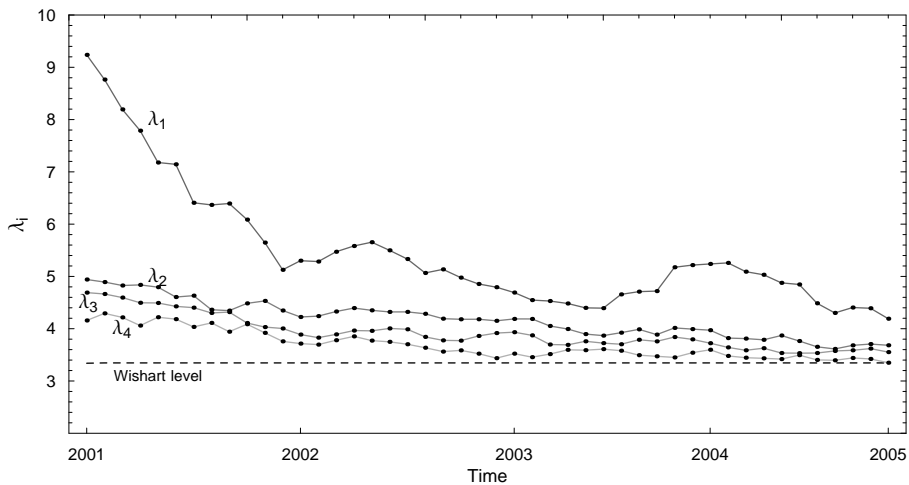}
\caption{Eigenvalues evolution of the sequence of $N \times
N=250\times250$ WIG20 correlation matrices translated with the
step of one month. The dashed line corresponds to noise level.}
\end{figure}

An optimal way to visualize the character of repeatable intraday
structures is to look at the eigensignals as defined by the
Eq.~(\ref{eq7}). They can be calculated for all the eigenvectors.
An explicit numerical verification confirms that they are
orthogonal indeed, i.e., the correlation matrix constructed out of
them is diagonal.
\begin{figure}[h!]
\hspace{-1cm} \epsfxsize 14.3cm \epsffile{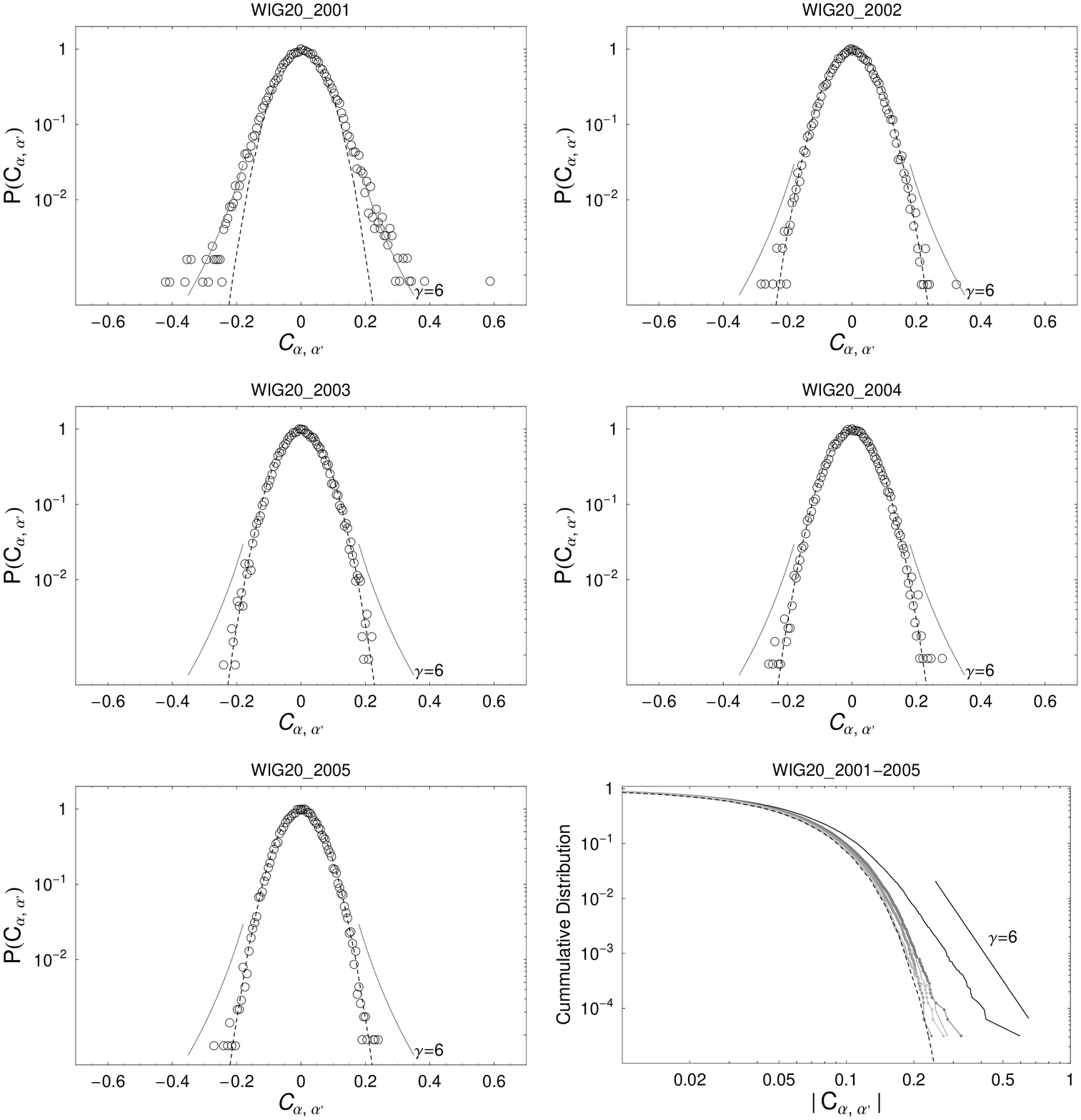}
\caption{Distribution of matrix elements $C_{\alpha,\alpha'}$ of
the $N_{\mathcal Y}\times N_{\mathcal Y}$ correlation matrices
$\bf C_{\mathcal{Y}}$ for the WIG20 variation during the intraday
trading time 10:01:30--16:10:00. The solid lines indicate the
power law fits to the tails of $\bf{C}_{2001}$ with the power
index $\gamma=6$ ($P(x)\sim x^{-\gamma}$). The dashed lines
corresponds to a Gaussian best fit.}
\end{figure}

The most relevant examples of such eigensignals - corresponding to
the two largest eigenvalues, for all the five years considered
here - are shown in Fig.~5. They both display a strong enhancement
of market activity just after the opening and lasting up to 60
min.
\begin{figure}[h!]
\hspace{-2.9cm} \epsfxsize 18.3cm \epsffile{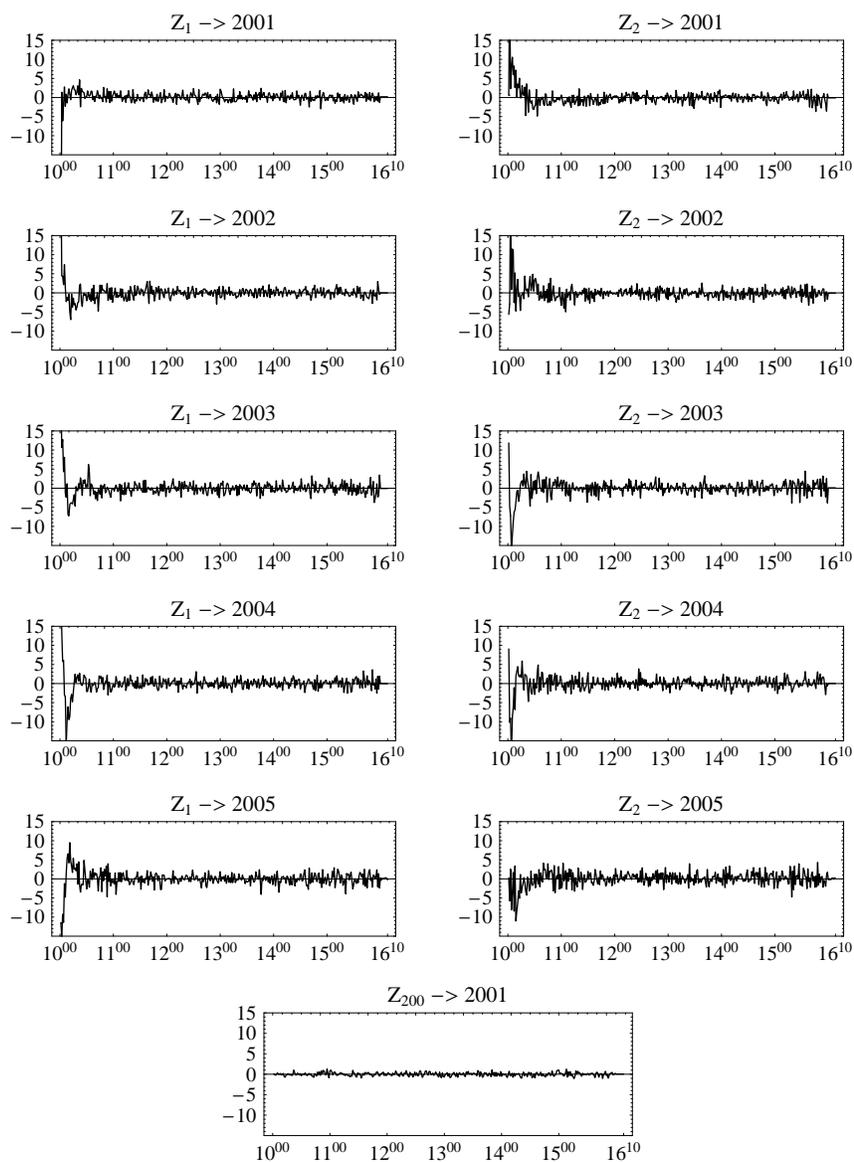}
\caption{Intraday eigensignals corresponding to the two largest
eigenvalues ($\lambda_1$, $\lambda_2$) calculated for five
calender years of WIG20(Warsaw Stock Exchange) index variation
during the intraday trading time 10:01:30--16:10:00. The last
graph is the same intraday  eigensignal but corresponding to
$\lambda_{200}$ for the year 2001.}
\end{figure}

Interestingly, an analogous enhancement before closing as
observed~\cite{Kwapien2} for the other markets, in case of the
WIG20 can be seen only rudimentary. For comparison one typical
eigensignal $(z_{200}(t_j))$ corresponding to the bulk of
eigenspectrum is shown in the bottom of Fig.~5. Its amplitude of
oscillations can be seen to be about an order of magnitude smaller
than for the previous leading eigensignals.

\subsection{Correlations among trading weeks}

Ability of the above formalism to detect and decompose some
potential repeatable structures in the financial patterns prompts
a question of correlations among different trading weeks. Our
WIG20 data set (Fig.~1) comprises $N=207$ full trading weeks and
thus allows to construct a $207 \times 207$ correlation matrix.
The lengths $T$ of the corresponding time series of 1 min returns
between monday opening and friday closing equals 1840. The upper
panel of Fig.~6 shows the resulting spectrum of eigenvalues.
\begin{figure}[h!]
\epsfxsize 7.5cm \hspace{0.5cm} \epsffile{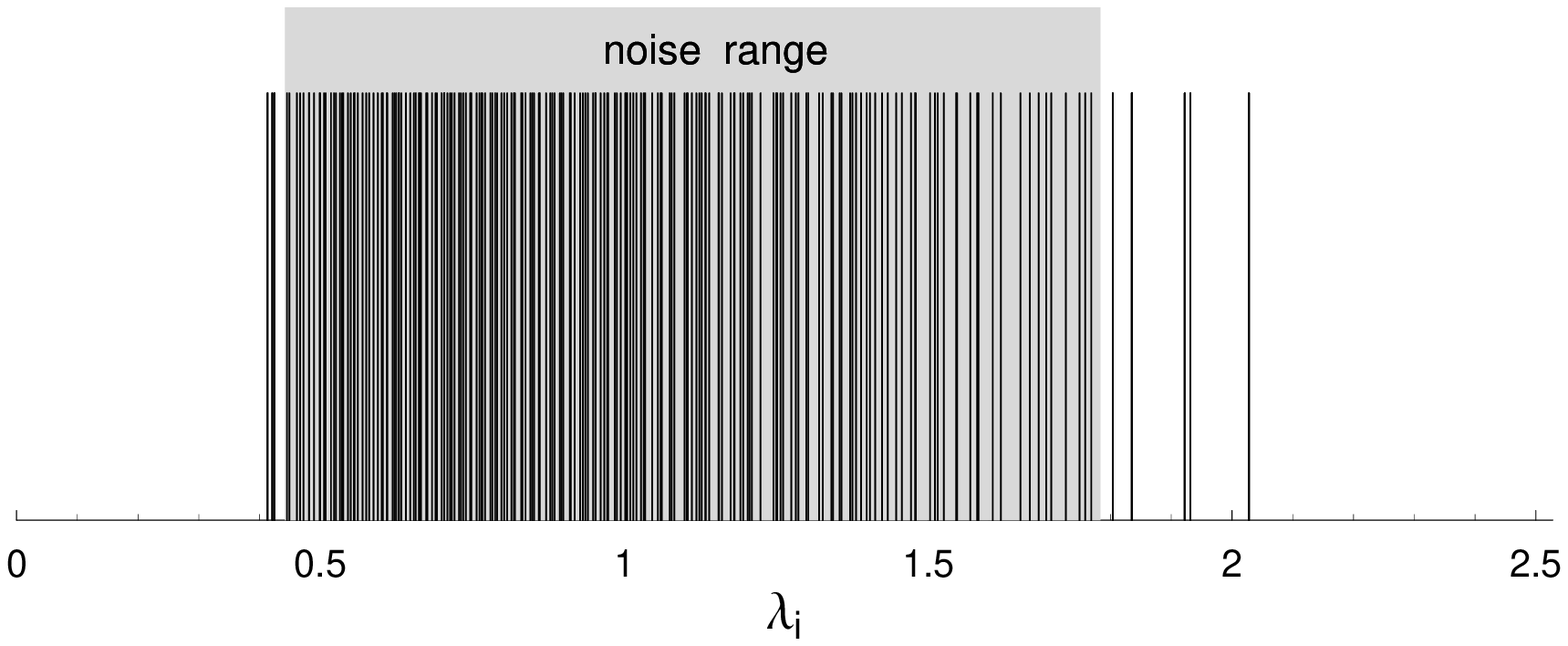} \vspace{1cm}
\epsfxsize 12.5cm \hspace{0cm} \epsffile{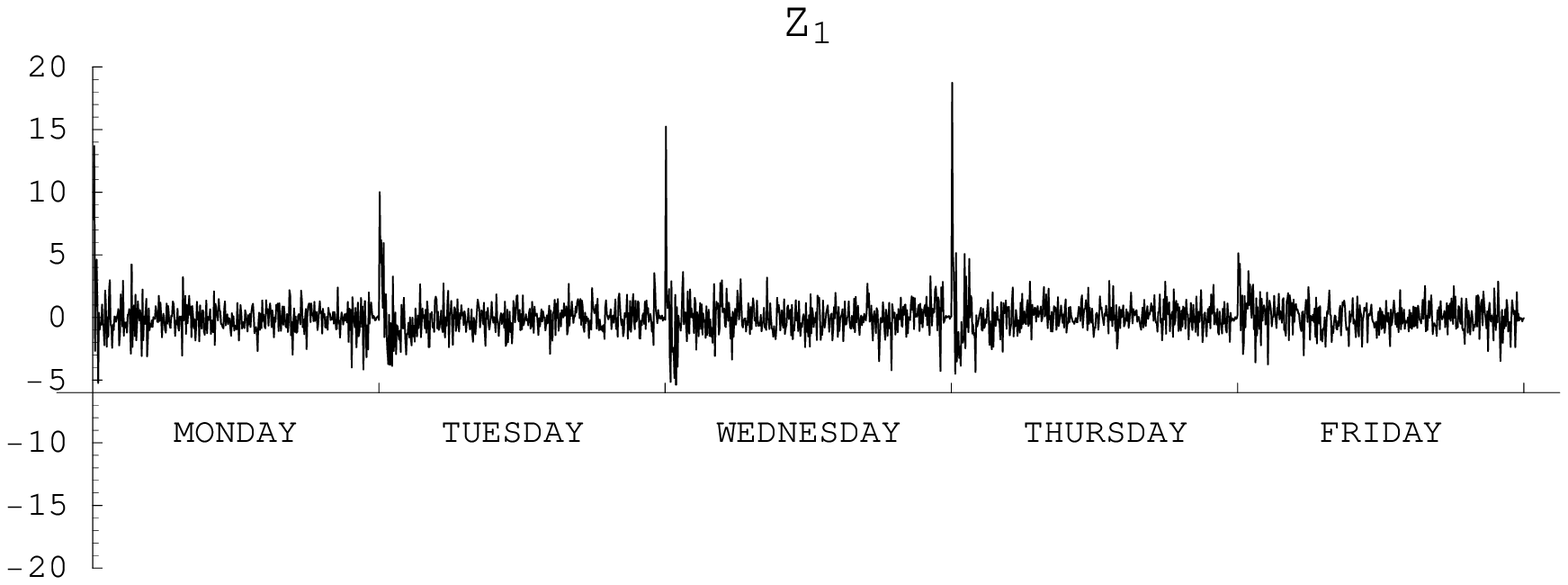}
\caption{$Top$ - Empirical eigenvalue spectrum of the $207\times
207$ correlation matrix calculated among the weekly time intervals
for the whole period of the WIG20 recordings as shown in Fig.~1.
The corresponding noise range of a random Wishart matrix with
$Q=1840/207$ ( $\lambda_{max}\approx 1.78$ and
$\lambda_{min}\approx 0.44$) is marked by the shaded field.
$Bottom$ - Intraweek (monday 10:01:30 -- friday 16:10:00)
eigensignal associated with the largest eigenvalue $\lambda_1$.}
\end{figure}
Most of them fall into the Wishart random matrix spectrum range
(shaded area) but at least three eigenvalues of our empirical
correlation matrix stay apart.
\begin{figure}[h!]
\hspace{2cm} \epsfxsize 7.5cm \epsffile{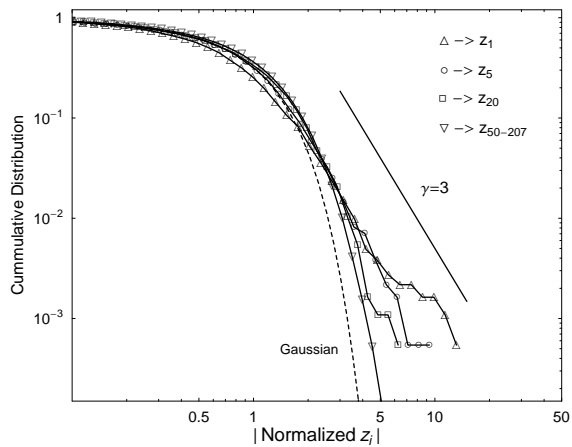}
\caption{Cumulative distributions of the moduli of normalised
intraweek eigensignals. Dashed line corresponds to a Gaussian
distribution and the solid line indicates a slope corresponding to
the inverse cubic power-law.}
\end{figure}
The eigensignal associated with the largest eigenvalue
$(\lambda_1)$ is shown in the lower panel of this Figure. It shows
an enhanced market activity at the connections between the days.
Interestingly however this activity is much stronger in the middle
of the week than in its beginning or in the end.\\
\indent Finally, such a decomposition of financial fluctuations
allows an instructive insight into the statistics of returns
distributions. This in itself constitutes one of the central
issues of econophysics. The related well identified stylized fact
is the so-called inverse cubic power-law~\cite{Gopi,Drozdz3,Rak}.
There exist also some consistency arguments that favor this
law~\cite{Gabaix}. Fig.~7 shows the cumulative return
distributions associated with several weekly eigensignals. Those
distributions that originate from the bulk of eigenspectrum can be
seen not to deviate much from a Gaussian even for such a short
time lag of 1 min. The fatter tails result from the fluctuations
filtered out by the eigensignals connected with the largest
eigenvalues. In particular, the most extreme events can be seen in
the first eigensignal, the one whose main component constitute
fluctuations commonly considered a daily or a weekly trend.

\section{Summary} The way of using the correlation matrix
formalism, as presented here, opens a promissing novel window to
view the character of financial fluctuations. In particular the
related concept of eigensignals allows to filter out all
repeatable synchronous patterns in the market daily or weekly
activity. They are connected with a few largest eigenvalues of the
corresponding correlation matrix. It is those eigensignals that
appear to be responsible for the fat tails in the return
distributions. The overwhelming rest of the spectrum stays within
the borders prescribed by the random ensemble of Wishart matrices
and fluctuations of the corresponding eigensignals are essentially
of the Gaussian type. As far as the WIG20 dynamics is concerned it
is interesting to notice a gradual weakening of the daily trend
effects when going from 2001 to 2005. A question remains whether
this effect is characteristic to this specific market or it takes
place in the other markets as well.

\end{document}